\documentclass[twocolumn]{aa}
\usepackage{epsfig}
\def\simlt{\ \raise -2.truept\hbox{\rlap{\hbox{$\sim$}}\raise5.truept   %
\hbox{$<$}\ }}
\def\simgt{\ \raise -2.truept\hbox{\rlap{\hbox{$\sim$}}\raise5.truept   %
\hbox{$>$}\ }}                                                          %

\def\be{\begin{equation}}
\def\ee{\end{equation}}
\def\newline{\hfil\break}

\def\la{\mathrel{\hbox{\rlap{\hbox{\lower4pt\hbox{$\sim$}}}\hbox{$<$}}}}
\def\ga{\mathrel{\hbox{\rlap{\hbox{\lower4pt\hbox{$\sim$}}}\hbox{$>$}}}}

\begin{document}

\title{SZ effect from Dark Matter annihilation}

   \author{S. Colafrancesco }

   \offprints{S. Colafrancesco}

\institute{   INAF - Osservatorio Astronomico di Roma
              via Frascati 33, I-00040 Monteporzio, Italy.
              Email: cola@mporzio.astro.it
%
             }

\date{Received / Accepted }

\authorrunning {S. Colafrancesco et al.}

\titlerunning {SZ effect from Dark Matter annihilation}

\abstract{We derive in this Letter the SZ effect induced by the secondary electrons produced in the
annihilation of Weakly Interacting Massive Particles (assumed here to be neutralinos) in
gravitationally bound structures dominated by Cold Dark Matter (CDM). We show that the DM induced SZ
effect has a specific spectral shape and an amplitude which increases for decreasing neutralino mass
$M_{\chi}$. The available SZ data on the Coma cluster set an upper limit on the quantity $\langle
\sigma V \rangle_A n^{2}_{\chi}$ which can be combined with the WMAP constraints on $\Omega_m h^2$ to
restrict the available neutralino models in the $\langle \sigma V \rangle_A - M_{\chi}$ plane. We
delineate various potential applications of this method to constraint the physical properties of the
Dark Matter particles from the study of galaxy clusters and dwarf spheroidal galaxies in the light of
the next coming high-sensitivity SZ experiments.

}

 \maketitle


\section{Introduction}

Dark Matter (DM) annihilation in the halo of galaxies and galaxy clusters have relevant astrophysical
implications. In fact, if DM is constituted by weakly interacting massive particles (WIMPs), their
annihilation can produce direct and indirect signals such as observable fluxes of positrons (e.g.,
Silk \& Srednicki 1984, Kamionkowski \& Turner 1991, Baltz \& Edsjo 1999) antiprotons (e.g., Bottino
et al. 1998), gamma rays (e.g., Chardonnet et al. 1995, Colafrancesco \& Mele 2001), neutrinos  (e.g.,
Gondolo \& Silk 1999, Hooper \& Silk 2004), radio emission (e.g., Colafrancesco \& Mele 2001), heating
of the hot intra-cluster gas (Totani 2004, Colafrancesco 2004). We do not have yet however, at
present, a definite detection of these emission features originating from DM annihilation.

As an alternative strategy, we explore here the consequences of the Compton scattering between the
secondary electrons produced from the WIMP anihilation in massive DM halos, like galaxy clusters, and
the CMB photon field.

Galaxy clusters are gravitationally dominated by Cold Dark Matter for which the leading candidate is
the lightest supersymmetric (SUSY) particle, plausibly the neutralino $\chi$. Experimental and
theoretical considerations for having a cosmologically relevant neutralino DM lead to bound its mass
$M_{\chi}$ in the range between a few GeV to a few hundreds of GeV (e.g., Bottino et al. 2003,
Belanger 2003). The decays of neutralino annihilation products (fermions, bosons, etc.) yield, among
other particles, energetic electrons and positrons up to energies comparable to the neutralino mass.
Here we notice that these energetic electrons and positrons (hereafter we will refer to these
particles as electrons because their distinction is not essential for our purpouses) can interact with
the CMB photons and up-scatter them to higher frequencies producing a peculiar Sunyaev-Zel'dovich
(1972, 1980, hereafter SZ) effect with specific spectral and spatial features. In this Letter, we will
describe the specific feature of the SZ effect produced by DM annihilation, SZ$_{DM}$, in galaxy
clusters and we will discuss the possibility to disentangle such specific SZ$_{DM}$ effect from the
other sources of SZ effect which are present in the same structures. We will finally discuss the
future experimental prospects for the detection of the SZ$_{DM}$ effect in DM halos. The relevant
physical quantities are calculated using $H_0 = 70$ km s$^{-1}$ Mpc$^{-1}$ and a flat,
vacuum-dominated CDM  ($\Omega_m = 0.3, \Omega_{\Lambda}=0.7$) cosmological model.

\section{The SZ effect from DM annihilation}

The calculation of the secondary electron spectrum from $\chi \chi$ annihilation in galaxy clusters
has been already presented in details by Colafrancesco \& Mele (2001) and here we will only recall the
relevant steps which are necessary for the present purpouses.

\subsection{Neutralino annihilation in galaxy clusters}

Following Colafrancesco \& Mele (2001) we assume a spherical, uniform top-hat model for the collapse
of a cluster with an average DM density $\bar {\rho} = \Delta(\Omega_0,z) \rho_b ( 1 - f_g)$ where the
background density is $\rho_b = \Omega_m \rho_c$ and $\Delta(\Omega_0,z)$ is the non-linear density
contrast of the halo which virializes at redshift $z$. Here $f_g$ is the gas mass fraction.
Redistributing the total mass, $M={4 \pi \over 3} \bar{\rho} R_p^3$, found within the radius $R_p= p
r_c$ (expressed in terms of the cluster core radius $r_c$), according to a density profile $\rho(r) =
\rho_0 g(r)$, we find that the central total density is $\rho_0 = {\bar{\rho} \over 3} {R^3_p \over
I(R_p) } ~,$ where $I(R_p) = \int_0^{R_p} dr r^2 g(r)$. In our phenomenological approach, we consider
mainly the case a central cusp profile, as indicated by the available N-body simulations. The central
cusp model described by a profile:
\be
g(r) = \bigg({r \over r_c} \bigg)^{-\eta} \bigg(1+ {r \over r_c} \bigg)^{\eta - \xi}
 \ee
gives a central density $\rho_0 = {\bar{\rho} \over 3} {p^3 \over I(p,\eta, \xi) } ~,$ where
$I(p,\eta, \xi)= \int_0^p dx x^{2-\eta}(1+x)^{\eta-\xi}$. With $\eta = 1$ and $\xi = 3$, the central
cusp model corresponds to the ``universal density profile'' which Navarro, Frenk \& White (1997)
showed to be a good description of cluster DM halos in N-body simulations of CDM hierarchical
clustering. Assuming that the DM density scales like the total cluster density, the central DM number
density is:
\be
n_{\chi,0} = {\bar{n} \over 3} {p^3 \over I} \ee
 where
 \be
\bar{n} \approx 4.2 \cdot 10^{-5} ~{\rm cm}^{-3} ~ \Omega_{\chi} h^2 \bigg[{M_{\chi} \over 100 ~GeV}
\bigg]^{-1} \bigg[{\Delta(\Omega_0,z) \over 400} \bigg]
~.
 \ee
The $\chi$ annihilation rate in a DM halo is $ R = n_{\chi}(r) \langle \sigma V \rangle_A ~,$ where
$n_{\chi}(r) = n_{\chi,0} g(r)$ is the neutralino number density and $\langle \sigma V \rangle_A$ is
the $\chi \chi$ annihilation cross section averaged over a thermal velocity distribution at freeze-out
temperature. Although the $\chi \chi$ annihilation cross section is a non-trivial function of the mass
and physical composition of the neutralino, to our purpose it suffices to recall that the $\chi$ relic
density is approximately given by:
 \be
 \Omega_{\chi}h^2 \simeq \frac{3 \times 10^{-27}~{\rm cm}^{3}~{\rm s}^{-1}}
                  {\langle \sigma V \rangle_A }
 \label{eq.sigmav}
 \ee
(Jungman et al. 1996). Hence, for the $\chi \chi$  annihilation, we shall assume a total cross section
of $ \langle \sigma V \rangle_A \approx 2.6 \cdot 10^{-26} ~{\rm cm}^3 ~{\rm s}^{-1} $ to be
consistent with the value $\Omega_m h^2 \sim 0.116$ derived from WMAP (Bennett et al. 2003). For
values of $\Omega_m h^2$ in the range $0.085 - 0.152$, the annihilation cross section is fixed to
within a factor less than three. Detailed studies of the relic neutralino annihilation (Edsjo 1997)
show that the above value is well inside the allowed range predicted in supersymmetric theories for a
wide choice of masses and physical compositions of neutralinos that can be relevant as CDM candidates.
Enhancing (suppressing) the $\chi$ annihilation rate will have on our results the simple effect of
rescaling the final electron spectra by the same enhancement (suppression) factor.

Neutralinos which annihilate inside a DM halo produce quarks, leptons, vector bosons and Higgs bosons,
depending on their mass and physical composition. Electrons are then produced from the decay of the
final heavy fermions and bosons (monochromatic electrons, with energy about $M_{\chi}$, coming from
the direct channel $\chi\chi \to ee$, are in general much suppressed, Turner \& Wilczek 1990). The
different composition of the $\chi\chi$ annihilation final state will in general affect the form of
the final electron spectrum.
Analytical expressions for the e$^{\pm}$ spectrum has been given  by Colafrancesco \& Mele (2001) and
we refer to this paper for further details.

The time evolution of the electron spectrum is given by the transport equation:
 \be
{\partial n_e (E,r)\over \partial t} - {\partial  \over \partial E} \bigg[ n_e (E,r) b(E)\bigg] =
Q_e(E,r)
 \ee
where $n_e(E,r)$ is the equilibrium spectrum at distance $r$ from the cluster center for the electrons
with energy $E$. The source electron spectrum, $Q_e(E,r)$, rapidly reaches its  equilibrium
configuration mainly due to synchrotron and Inverse Compton Scattering losses at energies $E \simgt
150$ MeV and to Coulomb losses at smaller energies (e.g., Colafrancesco \& Mele 2001). Since these
energy losses are efficient in the cluster atmosphere and DM annihilation continuously refills the
electron spectrum, the population of high-energy electrons can be described by a stationary transport
equation
 \be
 - {\partial  \over \partial E} \bigg[ n_e (E,r) b_e(E)\bigg] = Q_e(E,r)
 \ee
from which the equilibrium spectrum can be calculated. Here, the function $b_e(E)$ gives the energy
loss per unit time at energy $E$
 \be
b_e(E) =
b_0(B_{\mu}) \bigg({E \over GeV} \bigg)^2 + b_{Coul} ~,
 \ee
where $b_0(B_{\mu})= (2.5\cdot 10^{-17} + 2.54 \cdot 10^{-18} B_{\mu}^2)$ and $b_{Coul}= 7\times
10^{-16} [n(r)/1 cm^{-3}]$, if $b_e$ is given in units of GeV/s.
The source spectra we derived (see Fig.1 in Colafrancesco \& Mele 2001) arise from an analytic
approximation of the exact shape of the electron spectrum that tries to cope with the details of the
quarks and leptons decays and of the hadronization of the decay products. Detailed electron spectra
can also be obtained by using state-of-the-art Monte Carlo simulations, although the analytical
approximations used here can resume the relevant aspects of more detailed studies.

\subsection{The SZ effect from $\chi-\chi$ annihilation products}

The generalized expression for the SZ  effect  which is valid in the Thomson limit
for a generic electron population in the relativistic limit and includes also the effects of multiple
scatterings and the combination with other electron population in the cluster atmospheres has been
derived by Colafrancesco et al. (2003). This approach is the one that should be properly used to
calculate the specific SZ$_{DM}$ effect induced by the secondary electrons produced by $\chi \chi$
annihilation. Here we do not repeat the description of the analytical technique and we refer to the
general analysis described in Colafrancesco et al. (2003). According to these results, the DM induced
spectral distortion can be written as
 \begin{equation}
\Delta I_{DM}(x)=2\frac{(k_B T_0)^3}{(hc)^2}y_{DM} ~\tilde{g}(x) ~,
\end{equation}
where $T_0$ is the CMB temperature and the Comptonization parameter $y_{DM}$ is given by
\begin{equation}
y_{DM}=\frac{\sigma_T}{m_e c^2}\int P_{DM} d\ell ~,
\end{equation}
in terms of the pressure $P_{DM}$ contributed by the secondary electrons produced by neutralino
annihilation.
The quantity $y_{DM} \propto \langle \sigma V \rangle_A  n^2_{\chi}$ and scales as $\propto \langle
\sigma V \rangle_A M^{-2}_{\chi}$, providing an increasing pressure $P_{DM}$ and optical depth
$\tau_{DM} = \sigma_T \int d \ell n_e$ for decreasing values of the neutralino mass $M_{\chi}$.
The function $\tilde{g}(x)$, with $x \equiv h \nu / k_B T_0$, for the DM produced secondary electron
population can be written as
\begin{equation}
\label{gnontermesatta} \tilde{g}(x)=\frac{m_e c^2}{\langle k_B T_e \rangle} \left\{ \frac{1}{\tau}
\left[\int_{-\infty}^{+\infty} i_0(xe^{-s}) P(s) ds- i_0(x)\right] \right\}
\end{equation}
in terms of the photon redistribution function $P(s)$ and of  $i_0(x) = 2 (k_B T_0)^3 / (h c)^2 \cdot
x^3/(e^x -1)$, where we defined the quantity
\begin{equation}
\label{temp.media} \langle k_B T_e \rangle \equiv \frac{\sigma_T}{\tau}\int P d\ell = \frac{\int P
d\ell}{\int n_e d\ell} = \int_0^\infty dp f_e(p) \frac{1}{3} p v(p) m_e c ~,
\end{equation}
(see Colafrancesco et al. 2003) which is the analogous of the average temperature for a thermal
population (for a thermal electron distribution $\langle k_B T_e \rangle = k_B T_e$ obtains, in fact).
The photon redistribution function $P(s)= \int dp f_e(p) P_s(s;p)$ with $s = ln(\nu'/\nu)$, in terms
of the CMB photon frequency increase factor $\nu' / \nu = {4 \over 3} \gamma^2 - {1 \over 3}$, depends
on the electron momentum ($p$) distribution, $f_e(p)$, produced by $\chi \chi$ annihilation.
%
%

\begin{figure}[tbp]
\begin{center}
 \epsfig{file=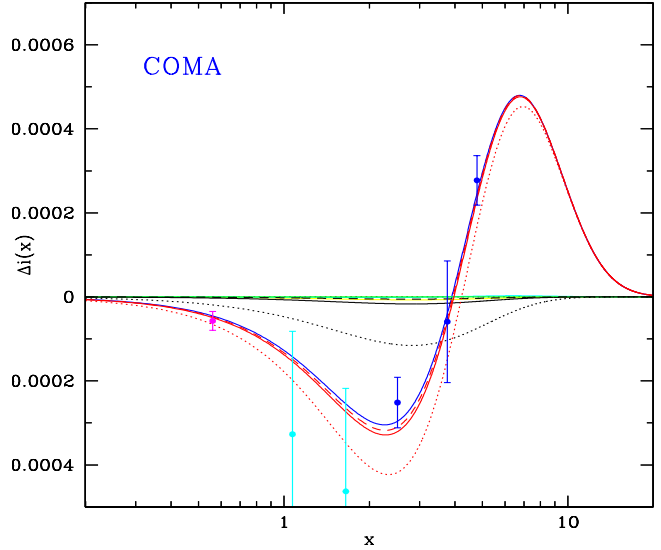,height=8.cm,width=9.cm,angle=0.0}
  \caption{\footnotesize{The overall SZ effect in Coma produced by the combination of various electron
  populations: thermal hot gas with $k_BT = 8.2 $ keV and $\tau = 4.9 \cdot 10^{-3}$ (solid blue curve)
  which best fits the available SZ data (DePetris et al. 2003); relativistic electrons which best fit
  the radio-halo spectrum (yellow curve) provide a small additional SZ effect (Colafrancesco et al. 2003);
  warm gas with $k_B T \approx 0.1$ keV and $n \approx 10^{-3} $ cm$^{-3}$ (cyan curve) provides a small SZ
  effect due to its low pressure (Colafrancesco 2004); DM produced secondary electrons with $M_{\chi} = 10$
  (black dotted curve), $20$ GeV (black solid curve) and $30$ GeV (dashed solid curve). A pure-gaugino
  $\chi$ reference model is assumed in the computations.
  The relative overall SZ effect is shown as the dotted, solid and dashed red curves,
  respectively. A zero peculiar velocity of Coma is assumed consistently with the available limits
  (Bernardi et al. 2002). SZ data are from OVRO (magenta), WMAP (cyan) and MITO (blue).
  }}\label{fig.sz_coma_tot_mchi102030}
\end{center}
\end{figure}

Fig.\ref{fig.sz_coma_tot_mchi102030} shows a comparison between the SZ effect obtained for a thermal
electron distribution with $k_B T = 8.2$ keV and that evaluated for the $\chi \chi$ annihilation for
different values of $M_{\chi}$. A major difference between the two spectral functions is the different
position of the zero of the SZ effect which is moved to higher frequencies in the case of the DM
produced electrons with respect to the case of the thermal distribution. As a consequence, the
SZ$_{DM}$ effect appears as a negative contribution to the overall SZ effect at all the frequencies
which are relevant for the SZ experiments, $x \sim 0.5 - 10$. Note that the amplitude of the SZ$_{DM}$
effect increases for decreasing values of the neutralino mass since the quantity $y_{DM} \propto
\langle \sigma V \rangle_A n^2_{\chi} \sim M^{-2}_{\chi}$ which accounts for the scaling $n_{\chi}
\sim M^{-1}_{\chi}$. A further $M^{-1}_{\chi}$ dependence is present in the electron source function
from $\chi \chi$ annihilation (see, e.g., Colafrancesco \& Mele 2001, their appendix).

\subsection{Application to galaxy clusters}

\begin{figure}[tbp]
\begin{center}
 \epsfig{file=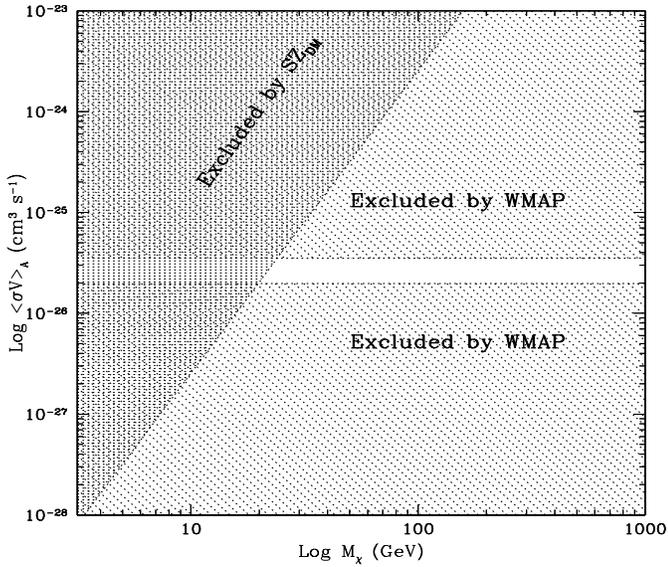,height=8.cm,width=9.cm,angle=0.0}
  \caption{\footnotesize{The constraints on the $\langle \sigma V \rangle_A$ -  $M_{\chi}$ plane
  set by the SZ effect from Coma. The heavily dashed area is excluded by the analysis of the SZ$_{DM}$
  in Coma. The SZ constraints are combined with the WMAP constraint $0.085
  \simlt \Omega_m h^2 \simlt 0.152$ (Bennett et al. 2003) which are translated on the quantity
  $\langle \sigma V \rangle_A$ through eq.(\ref{eq.sigmav}).
  }}\label{fig.sigmavmchilimits}
\end{center}
\end{figure}

The SZ effect in galaxy clusters is dominated by the Compton scattering produced by the hot, thermal
intra-cluster gas. However, it has been shown that there are additional sources of SZ effect which
might be produced by other electron populations of thermal (warm IC gas, see e.g. Colafrancesco 2004)
and non-thermal (relativistic electrons producing, e.g., the cluster radio-halo emission,
Colafrancesco et al. 2003, Colafrancesco 2004) origin residing in the cluster atmosphere. The SZ
effect unavoidably induced by DM annihilation in the same cluster combines with the previous SZ
effects to produce an overall SZ effect which, hence, retains a complex information about the physical
structure of the cluster atmosphere.
We will refer, in the following, to the case of the Coma cluster for which there are several SZ
observations available at different frequencies, from $\sim 30$ GHz to $\sim 350$ GHz. The Coma
cluster atmosphere is a complex environment which hosts various electronic populations: a hot, thermal
gas with $k_BT \approx 8.2 $ keV producing the bulk of the X-ray emission in the 1-10 keV region; a
population of relativistic electrons with a non-thermal spectrum, $n_{rel} \propto E^{-x}$ with $x
\approx 3.5 - 4$, and energy $E_e \approx  16.4 {\rm GeV} B^{-1/2}_{\mu} (\nu / {\rm GHz})^{1/2}$,
which are responsible for the radio-halo synchrotron emission (here $B_{\mu}$ is the magnetic field in
units of $\mu$G, see, e.g., Colafrancesco 2004 for a review); a possible warm gas halo with $k_B T
\approx 0.1$ keV claimed to be responsible for the observed soft X-ray/EUV emission excess (e.g.,
Bonamente et al. 2002, but see also Bowyer et al. 2004 for an alternative explanation); the secondary
electrons produced by the DM annihilation under the assumption that DM is mainly constituted by WIMPs.
The overall SZ effect from the combination of the different electronic populations in Coma can be
evaluated properly in the approach described by Colafrancesco et al. (2003).
The total SZ effect for values of $M_{\chi} = 10, 20$ and $30$ GeV is shown in
Fig.\ref{fig.sz_coma_tot_mchi102030}. To be consistent with the data, the SZ$_{DM}$ effect cannot be
due to neutralinos with $M_{\chi} \simlt 20$ GeV for the assumed value of  $\langle \sigma V
\rangle_A$. Since SZ$_{DM}$ depends on the quantity $\langle \sigma V \rangle_A n^2_{\chi}$, the
available SZ data set actually an upper limit in the $\langle \sigma V \rangle_A - M_{\chi}$ plane
(see Fig.\ref{fig.sigmavmchilimits}). Models with large values of $\langle \sigma V \rangle_A$ and low
values of $M_{\chi}$ which are found in the shaded area are excluded by the excess SZ$_{DM}$ effect in
Coma. The limits on $\langle \sigma V \rangle_A$ set by WMAP restrict further on the available region
of the $\langle \sigma V \rangle_A$-$M_{\chi}$ plane.

\section{Conclusions and future prospects}

We have shown  that the SZ effect induced by secondary electrons produced in $\chi \chi$ annihilation
is an unavoidable consequence of the presence and of the nature of Dark Matter in large-scale
structures. The analysis of the DM induced SZ effect in galaxy clusters provides a complementary probe
for the presence and for the nature of DM in cosmic structures. The available SZ observations on the
Coma cluster (see Fig.\ref{fig.sigmavmchilimits}) can already set a lower limit to the neutralino mass
of $M_{\chi} \simgt 17 - 20$ GeV ($M_{\chi} \simgt 13$ GeV at 90 \% c.l. with the adopted value of
$\langle \sigma V \rangle_A$), which are consistent with the limits set by accelerators (e.g.,
Belanger et al. 2003). The SZ$_{DM}$ signal does not strongly depend on the assumed DM density profile
at intermediate angular distances from the cluster center and on the DM clumpiness since $y_{DM}$ is
the integral of the total $P_{DM}$ along the line of sight.
The presence of a substantial SZ$_{DM}$ effect is likely to dominate the overall SZ signal at
frequencies $x\simgt 3.8-4.5$ providing a negative total SZ effect (see
Fig.\ref{fig.sz_coma_tot_mchi102030}). It is, however, necessary to stress that in such frequency
range there are other possible contributions to the SZ effect, like the kinematic effect and the
non-thermal effect which could provide additional biases (see, e.g., Colafrancesco et al. 2003).
Nonetheless, the peculiar spectral shape of the $SZ_{DM}$ effect is quite different from that of the
kinematic SZ effect and of the thermal SZ effect and this result allows to disentangle it from the
overall SZ signal.
An appropriate multifrequency analysis of the overall SZ effect based on observations performed on a
wide spectral range (from the radio to the sub-mm region) is required, in principle, to separate the
various SZ contributions and to provide an estimate of the DM induced SZ effect.
In fact, simultaneous SZ observations at $\sim 150$ GHz (where the SZ$_{DM}$ deepens the minimum with
respect to the dominant thermal SZ effect), at $\sim 220$ GHz (where the SZ$_{DM}$ dominates the
overall SZ effect and produces a negative signal instead of the expected $\approx$ null signal) and at
$\simgt 250$ GHz (where the still negative SZ$_{DM}$ decreases the overall SZ effect with respect to
the dominant thermal SZ effect) coupled with X-ray observations which determine the gas distribution
within the cluster (and hence the associated dominant thermal SZ effect) can separate the SZ$_{DM}$
from the overall SZ signal, and consequently, set constraints on the neutralino mass. Observations of
the radio-halo emission in the cluster can provide an estimate of the cosmic-ray electron population
and consequently an estimate of the associated non-thermal SZ effect (which is usually quite small and
with a different spectral shape at high frequencies, see e.g., Colafrancesco et al. 2003).
The high sensitivity planned for the future SZ experiments, especially at frequencies $x \approx 2.5$
and $x \simgt 3.8$, where the SZ$_{DM}$ more clearly manifests itself, can provide much stringent
limits to the additional SZ effect induced by DM annihilation. In this context, the next coming
PLANCK-HFI experiment has enough sensitivity to probe in details the contributions of various SZ
effects in the frequency range $x \approx 2 - 5$.
Because the amplitude of the SZ$_{DM}$ effect increases with decreasing values of $M_{\chi}$, the
high-sensitivity SZ experiments have - hence - the possibility to set reliable constraints to the
nature, amount and spatial distribution of DM in galaxy clusters. We will present elsewhere
(Colafrancesco 2004, in preparation) a more extended analysis of DM models in the context of SZ
observations.

An exciting possibility in this context could be offered by nearby (with small or zero peculiar
velocity) systems which are gravitationally dominated by Dark Matter, which contain little or no gas
(in either hot or warm forms) and show absence of non-thermal phenomena connected with the presence of
cosmic rays. In such ideal DM systems, the major source of SZ effect would be just the one due to the
annihilation of the WIMPs.  Systems which could be assimilable to the ideal "pure" DM halos are dwarf
spheroidal galaxies and/or low surface brightness galaxies. These systems seem to be ideal sites for
studying the DM annihilation indirect signals which reveal themselves in a variety of astrophysical
phenomena, whose main imprint is the specific gamma-ray emission (see, e.g., Evans et al. 2003). In
such a context, the possible detection of the DM induced SZ effect will provide an important
complementary approach which can be studied by more traditional astronomical techniques.

\begin{acknowledgements}
The author thanks the Referee for useful comments and P. Marchegiani and S. Profumo for useful
discussions.
\end{acknowledgements}


\begin{thebibliography}{}

\bibitem{} Baltz, E.A. \& Edsjo, J. 1999 Phys.Rev. D, 59,2,15
\bibitem{} Belanger, G. et al. 2003, preprint astro-ph/0310037
\bibitem{} Bennett, C.L. et al. 2003, ApJS, 148, 175
\bibitem{} Bernardi, M. et al. 2002, AJ, 123, 2990
\bibitem{} Bonamente, M. et al. 2002, ApJ, 585, 722
\bibitem{} Bottino, S. et al. 1998, Phys.Rev.D, 58, 1215
\bibitem{} Bottino, S: et al. 2003, Phys. Rev. D, 68, 043506
\bibitem{} Bowyer, S. et al. 2004, ApJ submitted, preprint astro-ph/0403181
\bibitem{} Chardonnet, P. et al.  1995, ApJ, 454, 754
\bibitem{} Colafrancesco, S. \& Mele, B. 2001, ApJ, 562, 24
\bibitem{} Colafrancesco, S. 2004, Journal of Nuclear Phys., in press
\bibitem{} Colafrancesco, S. et al. 2003, A\&A, 397, 27
\bibitem{} Colafrancesco, S. 2004, in 'Soft X-ray excess emission from clusters of galaxies',
R. Lieu \& J. Mittaz Eds., p.137-146 and p.147-154 (astro-ph/0403404)
\bibitem{} De Petris, M. et al., 2002, ApJ, 574, L119
\bibitem{} Evans, N.W., Ferrer, F. \& Sarkar, S. 2003, Phys. Rev. D in press (preprint astro-ph/0311145)
\bibitem{} Gondolo, P. \& Silk, J. 1999, Phys.Rev.Lett., 83, 1719
\bibitem{} Hooper, D. \& Silk, J. 2004, New Journal of Physics, 6, 023
\bibitem{} Itoh, N., Kohyama, Y. \& Nozawa, S. 1998, ApJ, 502, 7
\bibitem{} Jungman, G., Kamionkowski, M. \& Griest, K. 1996, Phys.Rep., 267, 195
\bibitem{} Kamionkowski, M. \& Turner, M.S. 1991 Phys.Rev. D, 43, 1774
\bibitem{} Navarro, J., Frenk, C. \&  White, S.D.M. 1997  ApJ, 490, 493
\bibitem{} Silk, J. \& Srednicki, M. 1984, Phys.Rev.Lett., 53, 624
\bibitem{} Sunyaev, R.A. and Zel'dovich, Ya.B. 1972, Comments Astrophys.
Space Sci., 4, 173
\bibitem{} Sunyaev, R.A. and Zel'dovich, Ya.B. 1980, ARA\&A, 18, 537
\bibitem{} Totani, T. 2003, Phys.Rev.Lett. in press, preprint astro-ph/0401140
\bibitem{} Turner, M.S. \& Wilczek, F. 1990, Phys.Rev.D, 42, 1001




\end{thebibliography}
\end{document}